# Self-Propelled Pedestrian Dynamics Model: Application to Passenger Movement and Infection Propagation in Airplanes

S. Namilae[1*], A. Srinivasan[2], A. Mubayi[3, 4], M. Scotch[5, 6] and R. Pahle[7]

[1]Aerospace Engineering, Embry-Riddle Aeronautical University, Daytona Beach, FL, USA,

[2] Department of Computer Science, Florida State University, Tallahassee, FL, USA,

[3] S.A.L Mathematical Computational Modeling Science Center, Arizona State University, Tempe, AZ, USA,

[4]School of Human Evolution and Social Change, Arizona State University, Tempe, AZ, USA

[5]Department of Biomedical Informatics, Arizona State University, Scottsdale, AZ, USA

[6]Center for Environmental Security, Arizona State University, Tempe, AZ. USA

[7]GeoData Center, Arizona State University, Tempe, AZ, USA

## Abstract

Reducing the number of contacts between passengers on an airplane can potentially curb the spread of infectious diseases. In this paper, a social force based pedestrian movement model is formulated and applied to evaluate the movement and contacts among passengers during boarding and deplaning of an airplane. Within the social force modeling framework, we introduce location dependence on the self-propelling momentum of pedestrian particles. The model parameters are varied over a large design space and the results are compared with experimental observations to validate the model. This model is then used to assess the different approaches to minimize passenger contacts during boarding and deplaning of airplanes. We find that smaller aircrafts are effective in reducing the contacts between passengers. Column wise deplaning and random boarding are found to be two strategies that reduced the number of contacts during passenger movement, and can potentially lower the likelihood of infection spread.

***** Assistant Professor of Aerospace Engineering, Corresponding Author, e-mail: namilaes@erau.edu

## 1. INTRODUCTION

The main factors determining whether or not transmission will successfully takes place in directly transmitted diseases are the ability of the agent to survive in the environment and/or the extent of the contact that occurs between infected and susceptible individuals of the host populations and their mobility within these populations. If a location has extremely high densities in a local area, it might pose high risks that may facilitate disease spread. That is, if the host and the agent are in close contact, the transmission of disease can be effected rapidly and easily. Thus, making it essential to estimate contacts for understanding disease dynamics. In 2003-2004 SARS outbreak, it was found that transmission rates fell during the epidemic, primarily as a result of reductions in population contact rates and improved hospital infection control [1]. Contact tracing of symptomatic infecteds have been found to be an effective control program because unidentified infecteds are most likely to be diagnosed under reasonable cost on resources [2]. The success of managing contacts and reducing drastically transmission rates have been directly seen in control of many diseases in the past such as smallpox [3], SARS epidemic [4], foot-and-mouth outbreak [5], and recent outbreak of Ebola [6].

Those who report more frequent social contact should be at a higher risk of infection during an epidemic, baring other considerations such as immunity and differences in susceptibility. There are number of large-scale empirical studies that attempt to estimate contact networks in case of sexually transmitted diseases [7]. However, relatively little effort has been devoted to infections spread by respiratory droplets or close contact. Instead, the contact structure for these infections has been assumed to follow a predetermined pattern governed by a small number of parameters that are then estimated using sero-epidemiological data [8] or using small non-representative populations survey [9]. Contact studies are especially relevant in high density and mobility areas such as in airports and airplanes. To address this lack of empirical knowledge, we develop a pedestrian dynamics model in the context of air-travel and robustly analyze the contact patterns.

There is direct evidence for spread of infection during commercial air-travel for many infectious diseases including influenza [10], SARS [11], tuberculosis [12], measles [13], norovirus [14] and malaria [15]. Models of infection transmission during air-travel [16-18] often utilize aggregate analysis based on the Wells-Riley equation [19] and do not account for discrete human interactions. Computationally intensive agent-based models (*e.g.* EpiSimdemics [20]) and stochastic models [21] include human interactions through behavioral rules. Such models are well suited for modeling simple interactions over large populations and geographical areas like entire urban areas [22]. Air travel however involves a high density of pedestrians over relatively small areas. Passengers move during boarding (ingress), deplaning (egress) and within cabin. Passengers otherwise not exposed to contagion may come into contact with contagion when they are in close proximity of infected passengers or contaminated surfaces during the high mobility phases of passenger entry and exit. Modeling the complete pedestrian trajectories and interactions as travelers move through airports and airplanes can help identify policies and procedures that reduce contacts between passengers and thereby reduce the infection spread.

Movement of passengers within an aircraft is a special case of a more general problem of pedestrian movement. This problem has been addressed using several approaches such as particle dynamics or social force models [23, 24], models based on cellular automata [25], fluid flow models [26], and queuing based models [27]. Social force models have specific advantages for studying passenger movement and contacts in airplanes. Each passenger is modeled individually and moves continuously; this enables individual trajectory evolution and estimation of the contacts between pedestrians.

Social force models of pedestrian movement are essentially based on molecular dynamics. In molecular dynamics, atoms are treated as Newtonian particles with forces between atoms described by interatomic potentials [28]. Social force models extend this concept to pedestrian movement. Here the forces are a measure of internal motivations of individual pedestrians to move towards their destination in presence of obstructions like other pedestrians and objects (e.g. chairs). Social force models have been applied to crowd simulations situations in panic [23], traffic dynamics [29], evacuation [30] and animal herding [31]. Algorithmic developments have included generation of force fields using visual analysis of crowd flows [32], explicit collision prediction [33], and collision avoidance [34].

One of the difficulties in modeling pedestrian movement is in estimating the parameters to be used in force fields. We address this problem by two approaches. Firstly, we formulate a local position based input to the self-propelling momentum of pedestrian particles. This modification to the equations of motion reduces the dependence on repulsive force-fields. Secondly, we use parallel computing in conjunction with available experimental data and vary the unknown model parameters over a vast design space to assess validated parameter combinations that explain the observed airplane exit data. We then use this pedestrian dynamics model to assess the optimal boarding and deplaning procedures that reduce contacts between individuals and can potentially reduce the infection spread. The simulations are performed on several airplane models and seating configurations with number of seats varying from 50 to 240.

## 2. PEDESTRIAN MOVEMENT MODEL FORMULATION

We model the motion of pedestrians using molecular dynamics based social force model [14]. The force $\overline{f}_i$ acting on *i^th^* pedestrian (or particle) can be defined as:

$$\overline{f}_i = \frac{m_i}{\tau}\left(\overline{v}_o^i(t) - \overline{v}^i(t)\right) + \sum_{j \neq i} \overline{f}_{ij}(t) \qquad (1)$$

where $\overline{v}_o^i(t)$ is the desired velocity of pedestrian and $\overline{v}^i(t)$ is the actual velocity, $m_i$ is the mass and $\tau$ is the time constant. The momentum generated by a pedestrian's intention results in a force that is balanced by a repulsion force $\overline{f}_{ij}(t)$. This force term represents the social forces wherein pedestrians avoid collisions with other people and objects. The dynamics of pedestrian movement is accomplished by obtaining the velocity $\overline{v}^i(t)$ and positions $\overline{r}^i(t)$ at next time steps as:

$$\overline{v}^i(t) = \frac{1}{m_i}\int \overline{f}_i dt \tag{2}$$

$$\overline{r}^i(t) = \int \overline{v}_i dt \tag{3}$$

We start with this basic model for airplane emergency evacuation [35]. In this paper, we modify these equations of motion by introducing a local neighbor dependence to the desired velocity $\overline{v}_o^i(t)$. In line forming applications like in an airplane entry or exit, the self-propelling force and desired velocity of $i^{th}$ pedestrian is dependent on the position of nearest pedestrian in the direction of motion. We make the following modification to the desired velocity of $i^{th}$ pedestrian $\overline{v}_o^i(t)$ in direction $\widehat{e}_1$.

$$\overline{v}_o^i(t).\widehat{e}_1 = \left(v_A + \gamma_i v_B\right)\left(1 - \frac{\delta}{\overline{r}_i\widehat{e}_1 - \overline{r}_k\widehat{e}_1}\right) \tag{4}$$

Here $\widehat{e}_1$ is the direction of desired motion. For example, for an exiting passenger in an airplane aisle, this would be the direction along the aisle. $\left(v_A + \gamma_i v_B\right)$ provides a distribution of desired speed for each pedestrian in the system. Here $\gamma_i$ is a random number, which varies for each pedestrian, enabling a distribution of speeds to account for differences in individual passengers. $\overline{r}_i$ and $\overline{r}_k$ denote the positions of $i^{th}$ and $k^{th}$ pedestrians where $k^{th}$ pedestrian is the nearest in $\widehat{e}_1$ direction and $\left(\overline{r}_i\widehat{e}_1 - \overline{r}_k\widehat{e}_1\right)$ would be the separation between them in direction $\widehat{e}_1$. In an airplane aisle the $k^{th}$ pedestrian would be the one directly in front of the $i^{th}$ pedestrian. $\delta$ is a distance constant such that at distance $\delta$ between $i^{th}$ and $k^{th}$ pedestrians the desired velocity of $i^{th}$ pedestrian is zero. When the distance between them is large the desired velocity of $i^{th}$ pedestrian is close to $\left(v_A + \gamma_i v_B\right)$ and reduces as they come close. This is representative of what happens in a line forming situation where pedestrians slow down as they get closer to persons in front of them. A similar feedback to self-propelling Langevin force has been applied to model the movement of ants [36].

The second part of particle dynamics in equation (1) corresponds to a repulsive social force term. Here we use repulsive part of the Lennard-Jones potential [19]. Most commonly monotonically varying exponential or power functions [23, 24, 30, 31, 33] have been used as potentials for calculating the force to represent repulsion of pedestrians to avoid collisions with other pedestrians and inanimate obstacles. There have been studies to obtain this force field using visual analysis [32].

The repulsive social force is also present in our model with the objective of ensuring pedestrian impenetrability. However, the balance between forces is not generated purely through repulsion; instead the desired velocity (and self-propelling force) also reduces when a pedestrian gets closer to another pedestrian who is in front of him. This is an advantage because there is adequate experimental data related to pedestrian speed measurements [37, 38], whereas determining an accurate repulsion force-field or potential is a relatively difficult task.

## 3. MODEL VALIDATION AND PARAMETER ESTIMATION

The formulation described above is implemented in a molecular dynamics code [41] and applied to the problem of airplane boarding and deplaning. The model parameters are varied over a vast design space and compared to real time observations of airplane deplaning to validate the models.

There are several parameters in the model such as maximum walking speed $(v_A + v_B)$, random variation $\gamma_i$, distance parameter $\delta$, two parameters for the Lennard-Jones repulsive force terms. In addition to pedestrian evolution through social force dynamics, we introduce behavioral aspects in the pedestrian movement. For example, in airplane boarding and deplaning we apply a time delay ( $t_{lug}$ ) for passengers to load cabin baggage and this is one of the varied parameters. Experimental data is available only for some of the parameters like walking speed range [37, 38]. Also observed exit times during deplaning can be found in the literature [39, 40] for a few commercial airplane models such as (a) Boeing 757-300 with 240 seats, (b) Boeing 757-200 with a single economy class and 201 seats (c) Boeing 757 with economy and first class with 182 seats (d) Airbus A320 with 144 seats and (e) CRJ200 with 50 seats.

We model the seating arrangements for these five airplanes for simulations in this paper. The seating arrangement is as shown in Figure 1. The passengers, seats and walls are all modeled as particles. The seat and wall particles can exert forces on pedestrian particles but only the pedestrian particles evolve in time according to equations 1 to 4. To assess the parameters for model validation we vary them over a wide design space and find values which can satisfy following requirements

(1) Same set of parameters should predict the observed deplaning times and average flow rates for all the five planes modeled.
(2) The model with correct parameters should be able to replicate, real time observations such as front to back unloading of airplane and clustering during line formation.

In addition, we limit the walking speed to ranges observed in literature [37, 38].

Variation of the model parameters over a wide design space requires large number of simulations. We utilize a parallel algorithm wherein each processor simulates pedestrian movement with a parameter combination. The parallel code is implemented on National Center for Supercomputing Applications' (NCSA) Bluewaters supercomputer. The resulting data is analyzed for deplaning time, flowrate, front to back unloading features to determine valid parameters. To reduce the number of simulations required, parameter evaluation is done in two passes. This first round narrows the parameter variations and second round is used for final estimation.

Figure 2 shows a modified parallel coordinate plot for the parameters varied in the first pass and second pass for simulations Airbus 320 deplaning. There are totally 144 seats, 24 First class and 120 economy in this seating arrangement and observed deplaning time range is 8 to 10 minutes. The parameter ranges that potentially result in these exit time and replicate observed behavior are broadly identified in the first pass and refined in the second pass. This procedure is repeated for all five airplane models considered in the study and parameter ranges that produce adequate results for all five airplanes are identified. Out of more than 40,000 parameter combinations we find a set of 26 parameter combinations (at different pedestrian speeds) that predict correct deplaning times and replicate the front to back deplaning behavior for all the models considered here. The comparison between simulation results with the parameter combinations and observed data is shown in Figure 4. The time evolution of particles during deplaning for airbus A320 airplane for one of the simulations is shown in Figure 3.

A value of 1.33 ft is obtained for the distance constant (at distance $\delta$ between two pedestrians, the desired speed of rear pedestrian is zero) from the parameter analysis. There are different Lennard-jones repulsion force parameters for different pedestrian desired speeds as shown in figure 2. The same algorithm can be used for determining the parameter ranges for any other form of the force field. Jakub and coworkers report that average walking speeds of men and women of different age ranges varies from 2.2 to 3.5 ft/s for slow walking and 3.5 to 5.1 ft/s for medium paced walking. Note that the speed in the plots is the maximum possible speed $\left(v_A + \gamma_i v_B\right)$ with no obstacles. The random variable $\gamma_i$ facilitates variation of speed for individual passengers. This is reduced when encountering another passenger or obstacle according to equation (4).

The same concepts are applied for modeling airplane boarding. The boarding time and trajectories depend significantly on the boarding strategy (for example front to back boarding, random boarding etc.). As a base line we use a random order for boarding of passengers and one set of validated input parameters from deplaning simulations. The aisle delay for boarding is increased to 20 seconds when the neighbors of target pedestrians are already seated. Figure 6 shows the time evolution of pedestrians for boarding for airbus A320 airplane seating configuration.

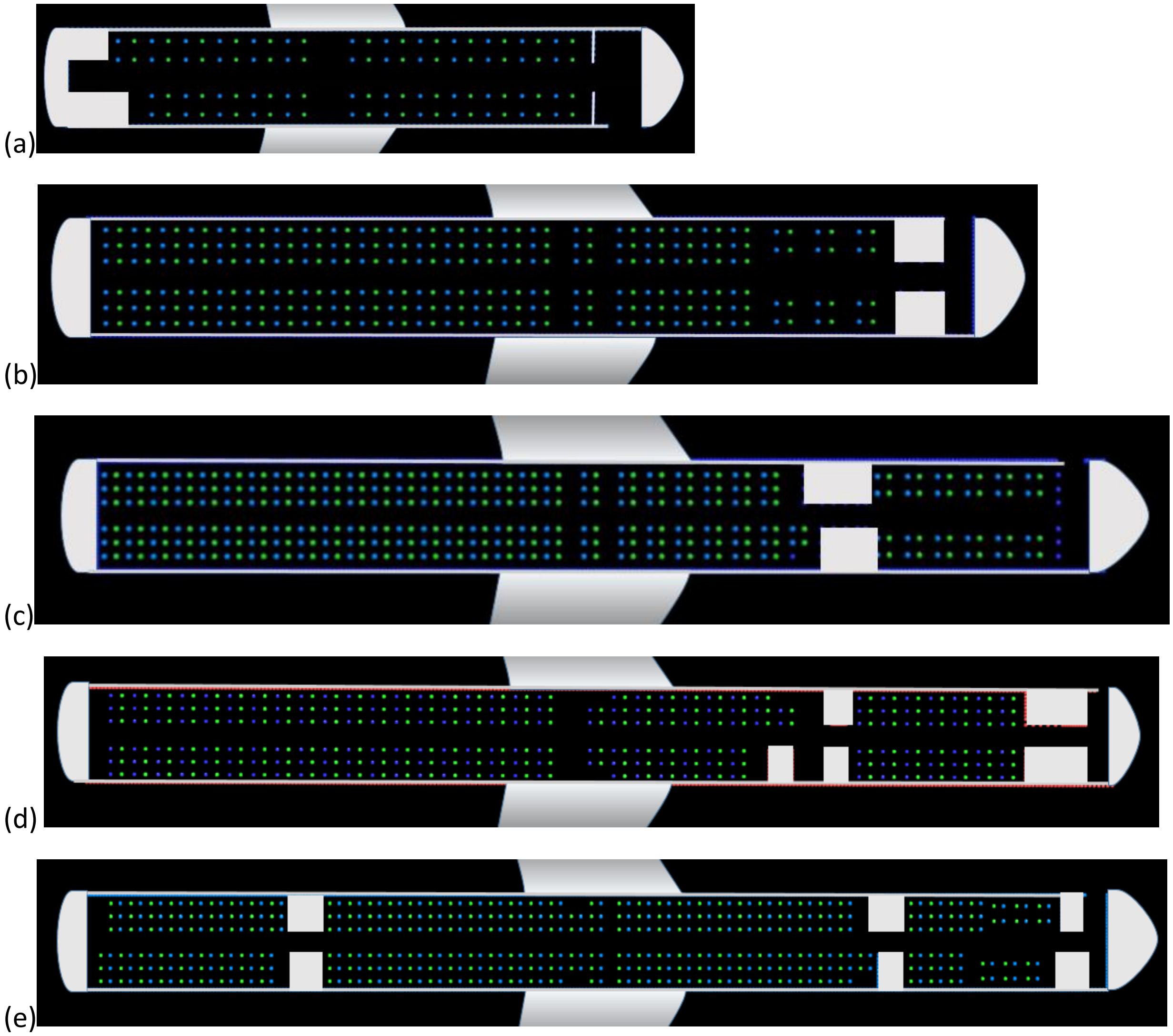

*Figure 1. Airplane configurations considered in the study (a) CRJ200 with 50 seats, (b)A320 with 144 seats (c( Boeing 757-200 with 182 seats, (d) Boeing 757-200 with 201 seats and € Boeing 757-300 with 240 seats.  Green dots represent pedestrian particles and blue dots represent fixed seat particles.*

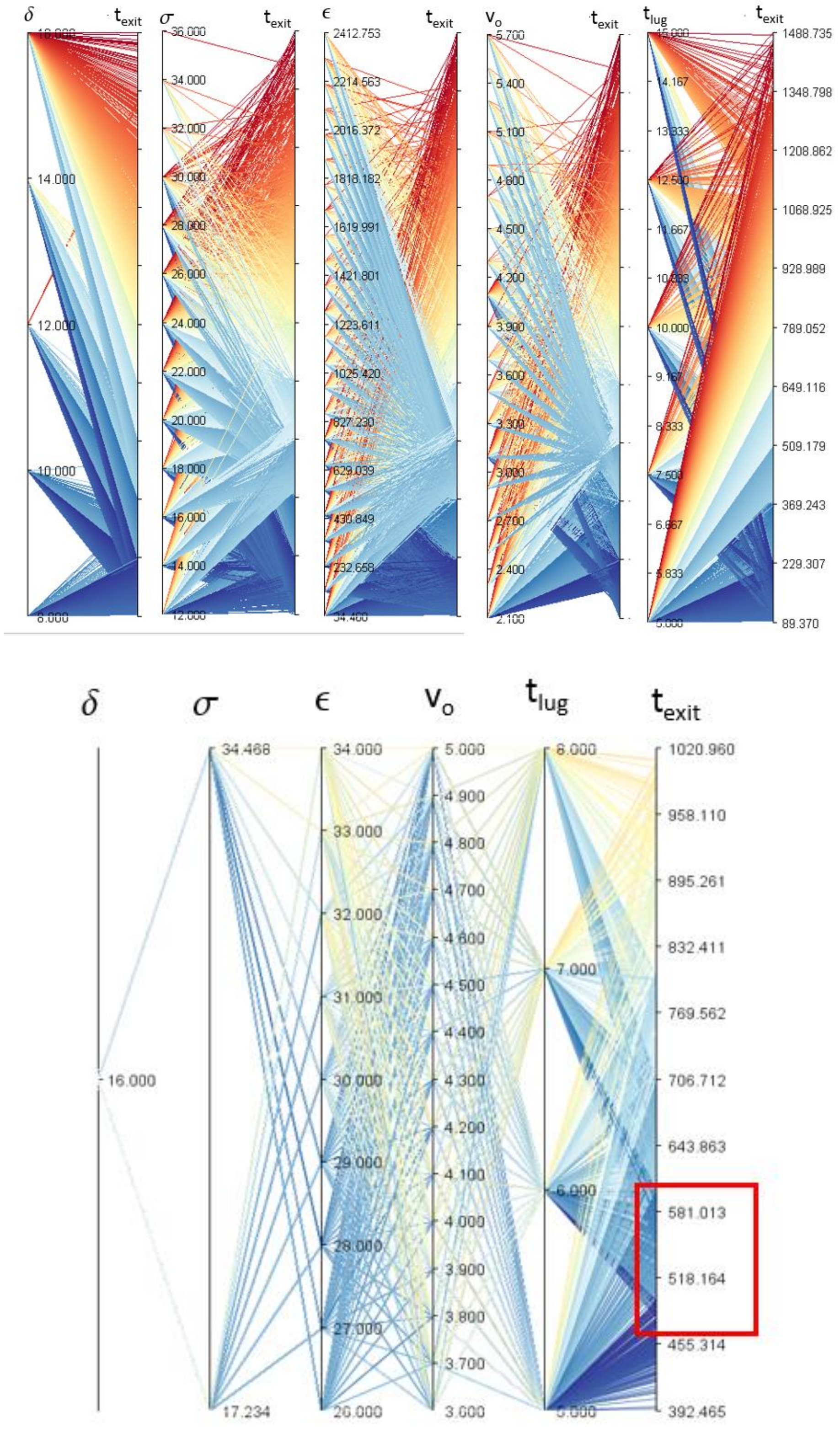

*Figure 2. Parallel coordinate plots show the variation of model parameters over two sets of simulations resulting in different exit times and trajectories.*

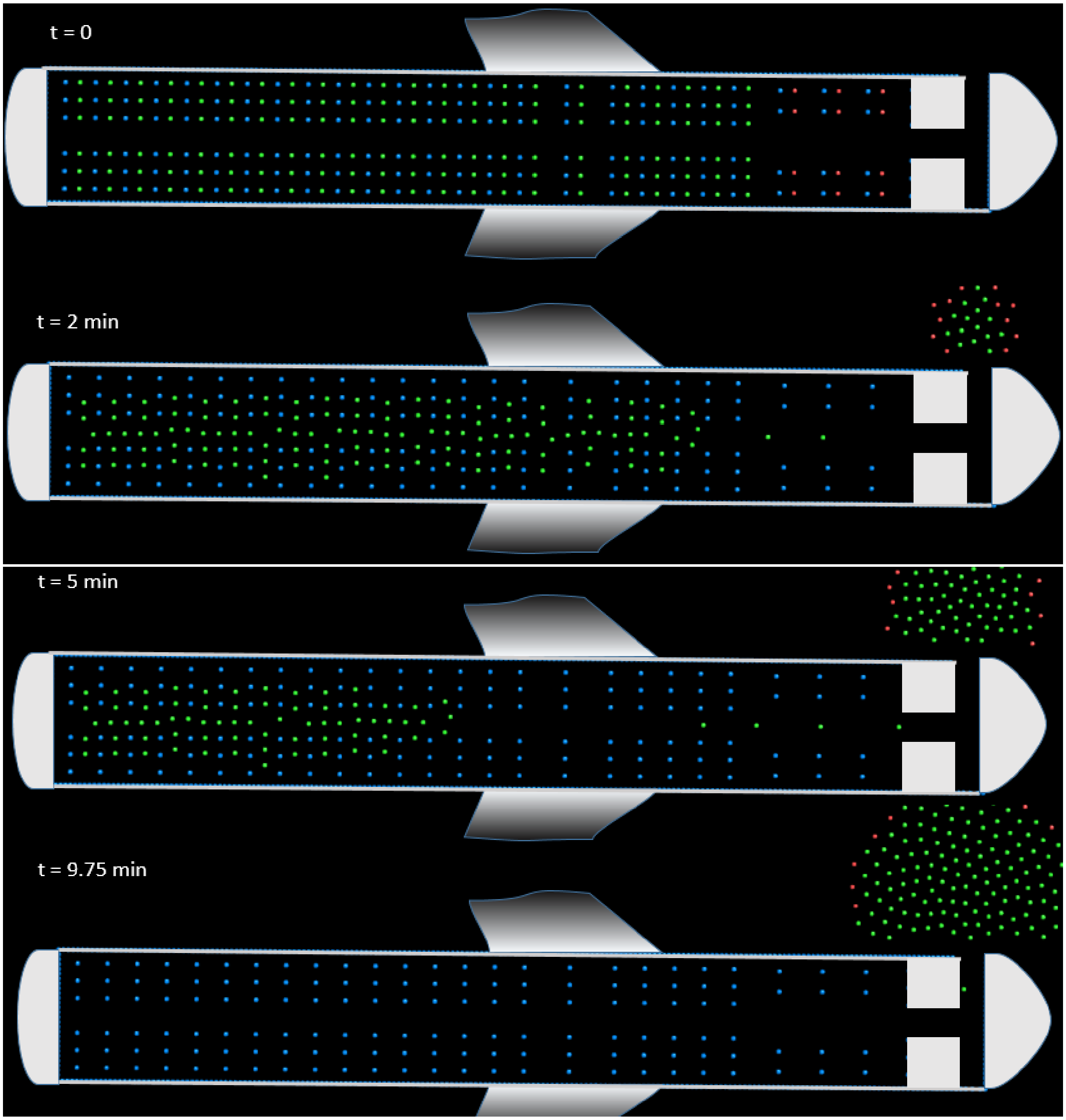

*Figure 3. Time evolution of pedestrians deplaning Airbus A320 airplane seating configuration with 144 seats.*

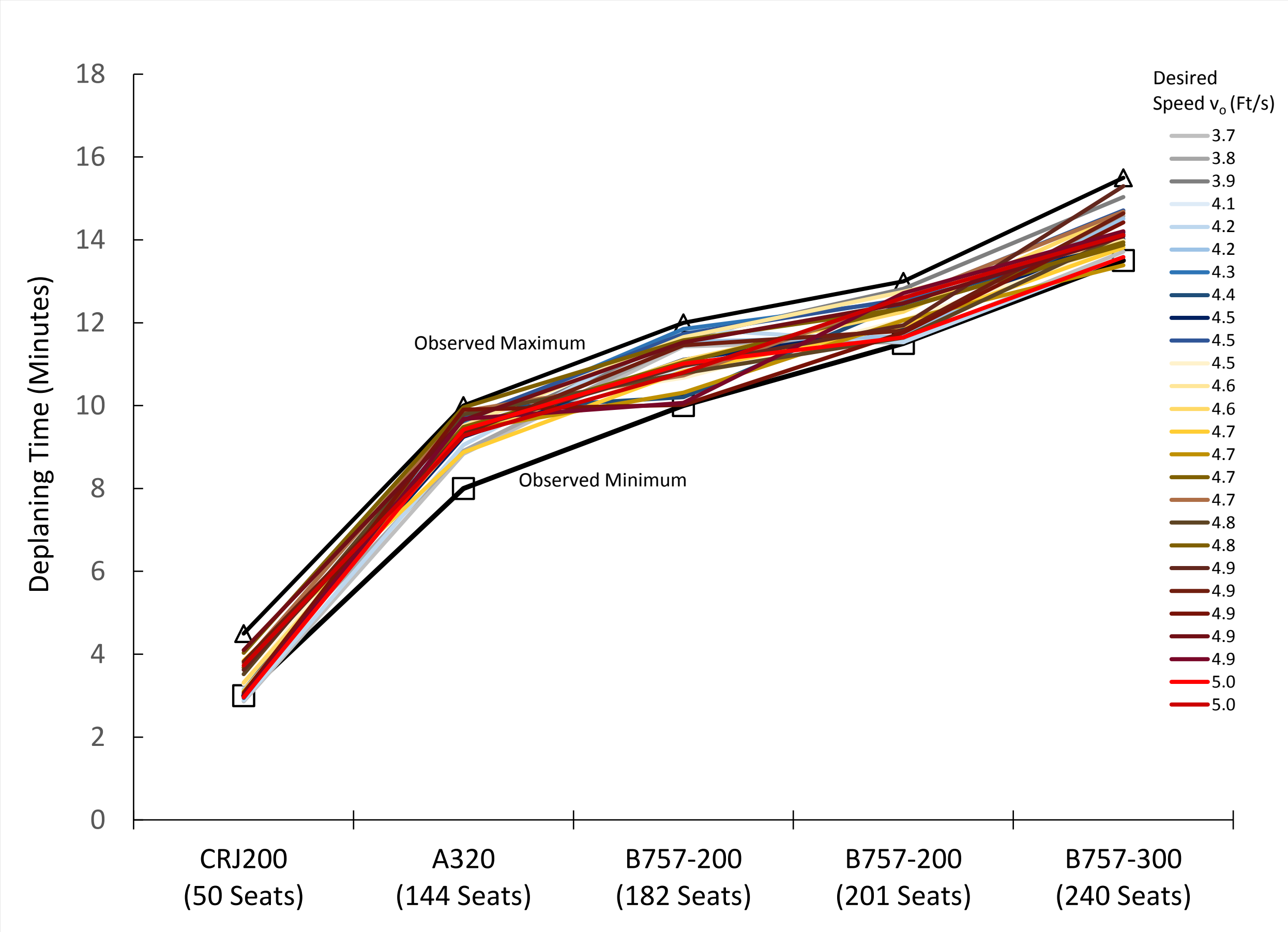

*Figure 4. Model parameter combinations that produce results that are comparable to observed data on airplane deplaning.*

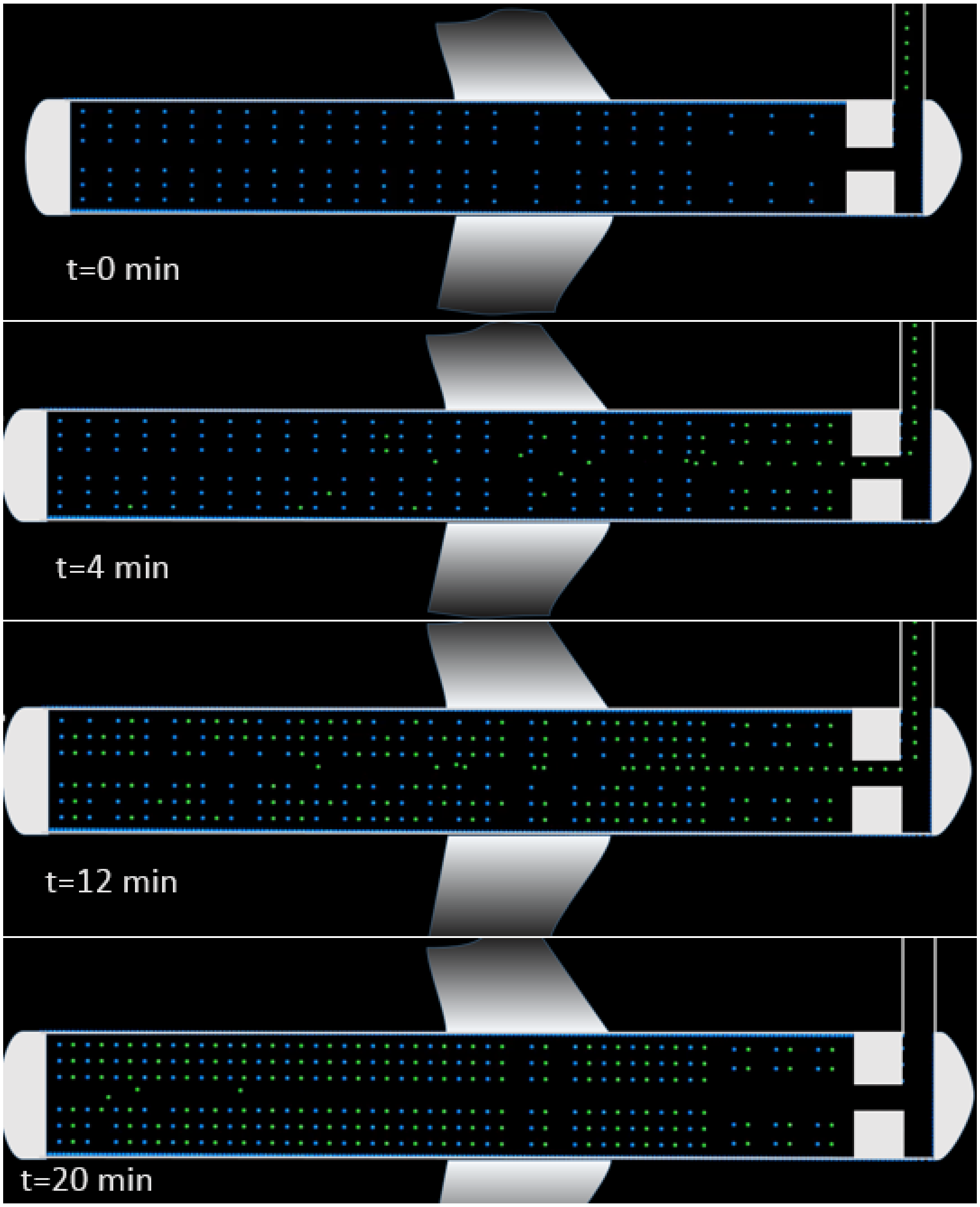

*Figure 5. Time evolution of pedestrians boarding Airbus A320 airplane seating configuration with 144 seats.*

## 4. RESULTS AND DISCUSSION

### *4.1. Number of contacts.*

Mangili and Gendreau [41] suggest that airborne transmission and large droplet transmission are the highest risk transmission mechanisms in aircraft cabin environment. In both the cases, passengers need to be in a certain proximity to the index case to be exposed to contagion. The likelihood of passengers coming into contact with each other and contaminated surfaces is higher during the high pedestrian mobility phases like boarding and deplaning. We now apply the pedestrian movement model to assess the number of contacts between passengers within a certain threshold distance. This information can be used to devise strategies to minimize such contacts and thereby reduce infection spread. The contacts are calculated by counting the number of times two pedestrian particles are within a threshold distance sampling the data every 1.25 seconds. Only the new contacts are counted in this process and the time for contact is more than 2.5 seconds, however time variation for contacts is not considered in detail here. Figure 6 shows the variation in total number of human-human contacts for the different airplanes considered for the validated parameter combinations from Figure 4. The variations for different parameter combinations are not that significant especially for smaller airplanes. This is because the number of contacts depends on the trajectory of movement of passengers which is similar for many parameter combinations.

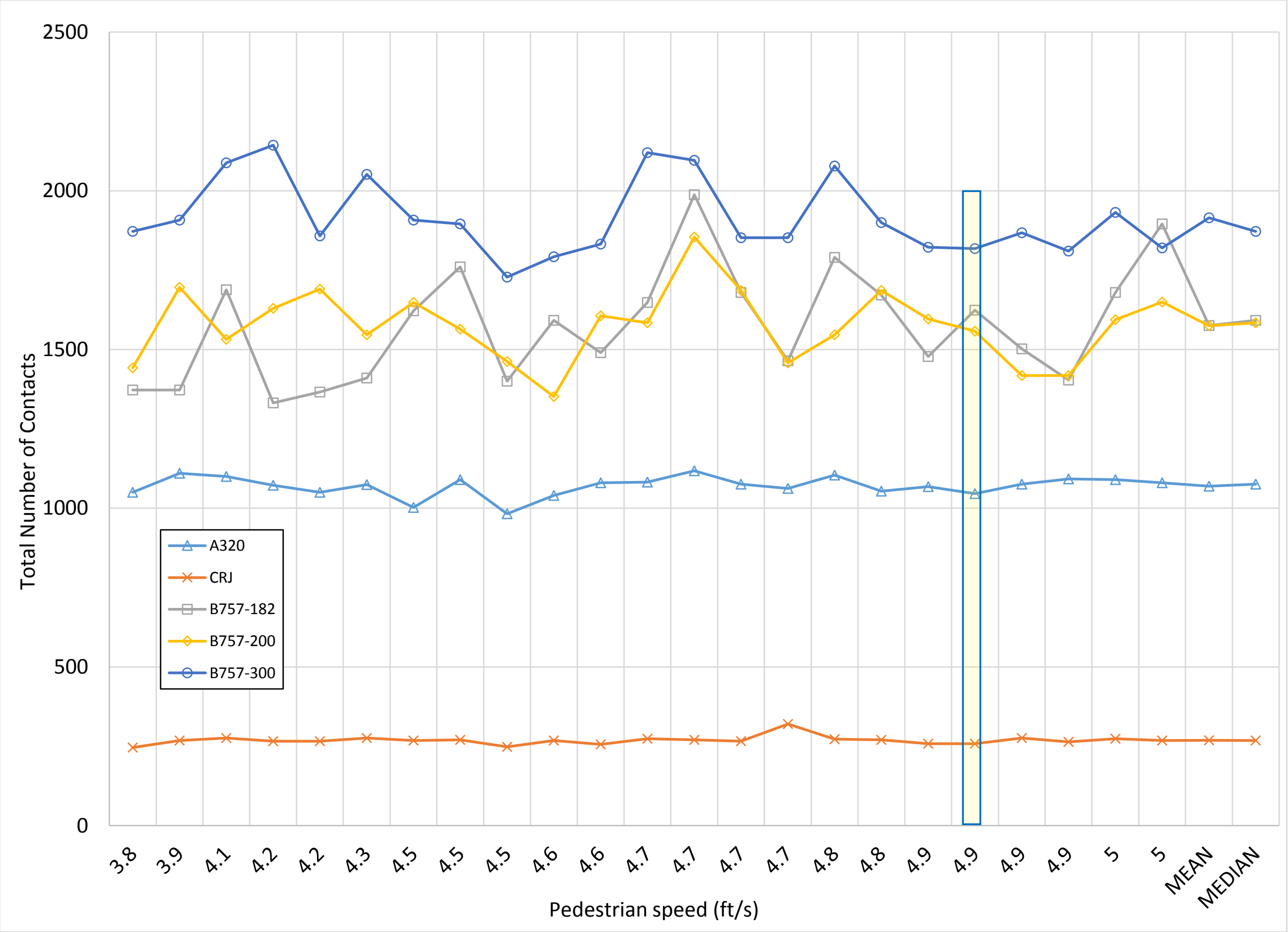

*Figure 6. Variation of Human-Human contacts with threshold distance of 18 inches for valid parameter combinations*

For further simulations we use the parameter combination highlighted in Figure 4. The total number of contacts is close to median for the planes considered with this combination of parameters. We now look at some details of airplane related policies that affect the number of human-human contacts. Policies that reduce the number of contacts can possibly be effective in reducing the propagation of infectious diseases during air-travel.

### *4.2. Comparison of Boarding and Deplaning:*

In Figure 7 and Figure 8 we show the number of contacts for the different airplanes for deplaning and boarding of aircraft respectively. The data is presented for two threshold distances of 18 inch and 30 inch. The 18 inch threshold distance between pedestrian particles indicates close proximity where pedestrians can potentially touch each other. The 30 inch threshold is presented for comparison at slightly larger distances. The default strategy is used for deplaning, i.e. there are no restrictions on which passengers exit first. This results in front to back exit as shown in Figure 3. For ingress this data corresponds to boarding in random order and is averaged over a hundred simulations. More people come into contact with each other during boarding than deplaning for all the seating configurations studied here. The increase is more pronounced for larger airplanes; for example for a 18 inch threshold, there is a 68 % increase to 341 contacts during boarding over the 202 contacts in deplaning for a 50 seater CRJ200, but the corresponding increase is 135% in the larger 240 seat Boeing 757-300 configuration. Also the difference between boarding and deplaning is much higher when the threshold distance for contact measurement is increased. For the 30 inch contact threshold boarding leads to about 3-5 times more contacts than deplaning for the different seating arrangements.

Figures 7 and 8 also show the comparison between the number of contacts for economy and first class passengers. Three of the airplane configurations considered, Airbus A320 (12 first class out of 144 seats), Boeing 757-200 (24 first class out of 182 seats) and Boeing 757-300 (12 first class out of 240 seats) have two classes while the other two configurations have a single class. Figure 9 shows the percentage of first class passenger contacts and first class seats for the airplanes considered. The number of contacts for first class seats is disproportionately lower compared to ratio of first class seats. The location of first class seats closer to the exit and the larger distance between the seats are obvious explanations for this difference.

The effect of size of airplane on the number of contacts and thereby infection spread can be assessed through these results. Figure 10 shows the number of contacts for transporting 1000 passengers using the different airplane models considered. These numbers include default boarding and deplaning methods on multiple flights with a particular airplane model to transport 1000 passengers. Smaller airplanes are more effective in reducing the number of contacts compared to larger airplanes, however, the advantage of airplane size reduces as airplane seating capacity increases.

### *4.3. Boarding and Deplaning strategies*

The strategy for boarding and deplaning of an airplane has an effect on the number of contacts between passengers. Moreover, it is one of the aspects of air travel that can be modified with relative ease if an infectious disease is prevalent. Boarding and deplaning strategies have been studied from the point of view of efficiency [27, 38, 39 and 42]. In our study, we used the pedestrian dynamics model to assess the number of contacts for different airplane entry and exit strategies.

In Figure 11, we show the number of contacts for deplaning strategies for the five airplanes studied here. The deplaning strategies are (a) baseline or default deplaning where there are no restrictions on which passengers exit first, (b) section wise deplaning, starting with first class followed by two sections for economy, (c) alternate columns, *i.e.* all aisle seats exit first followed by middle seats, and then window seats and (d) alternate rows, *i.e.* passengers in odd numbered seating rows first exit followed by even numbered seating rows. The baseline strategy is the most common deplaning approach. There is a distinct advantage in lowering the number of contacts if column wise exiting strategy is adopted compared to the baseline deplaning. This is an effective deplaning strategy for different contact thresholds and airplane models. In terms of exit efficiency, column wise deplaning is within two minutes of the default approach. Section-wise deplaning is very similar to default deplaning except the passengers in aft sections wait till all passengers in the section ahead of them exit, consequently there is a small reduction in number of contacts and an increase in exit time. Deplaning by alternate seat rows is the most ineffective and increases the contacts significantly. This is because pedestrian particles separated by a seat row come within contact threshold which would not have happened in other deplaning strategies.

In Figure 12, we show the number of contacts for different boarding strategies. Different airlines follow different boarding procedures to reduce the turn time at gates and this problem has been looked at quite extensively from operations research perspective [27, 39 and 42]. We considered different boarding strategies including (a) Boarding in a random order, (b) Section-wise boarding with only two sections one for first class and another for economy. Within the section, passengers board at random. (c) Section-wise boarding similar to (b), but with three sections. This is the most commonly used approach in zone-wise boarding employed by many airlines. (d) Boarding by columns with all window seat passengers boarding first followed by middle seats followed by aisle seats. Because of the random order of boarding either in entire airplane or within the sections, we averaged 100 simulations to determine the number of contacts for a particular boarding strategy (Figure 11). The number of contacts between pedestrian particles is clearly lowest for random boarding compared to any other ordered boarding approach. Boarding using multiple sections of airplane which is one of the common approaches currently followed results in a relatively high number of contacts. The entire set of pedestrians are ordered randomly for approach (a) while smaller sets of pedestrians belonging to specific sections, as in (c), or columns, as in (d), are randomly ordered in other approaches. This is one of the reasons for the increase in the number of contacts for (c) and (d). Comparing the boarding times, random boarding (a) takes up a higher boarding time than other approaches. For example in the 182 seat Boeing 757-200 there is a difference of seven minutes in boarding time between random boarding and boarding with three sections.

The contrast between approaches that optimize turn time versus those that reduce the number of contacts need to be considered in assessing air-travel procedures.

## 5. Summary

Mathematical models of infectious diseases transmission by the respiratory or close-contact route are increasingly being used to determine the impact of possible interventions. However, mixing patterns, known to be critical determinants for understanding disease dynamics from models, have little or no empirical basis [43, 44]. To better understand contact structure, a social force based pedestrian movement model with location dependence of self-propelling terms is formulated. The model parameters are validated by comparing with observed data on pedestrian movement in airplanes. In particular, this model is used to study the trajectories and contacts between people in several airplane configurations. We suggest a few air travel policies that can reduce the total number of contacts between passengers and thereby potentially reduce infectious disease spread.

**Acknowledgements**

The authors gratefully acknowledge the support of NSF-ACI award No. 1524972 (Simulation-Based Policy Analysis for Reducing Ebola Transmission Risk in Air Travel).

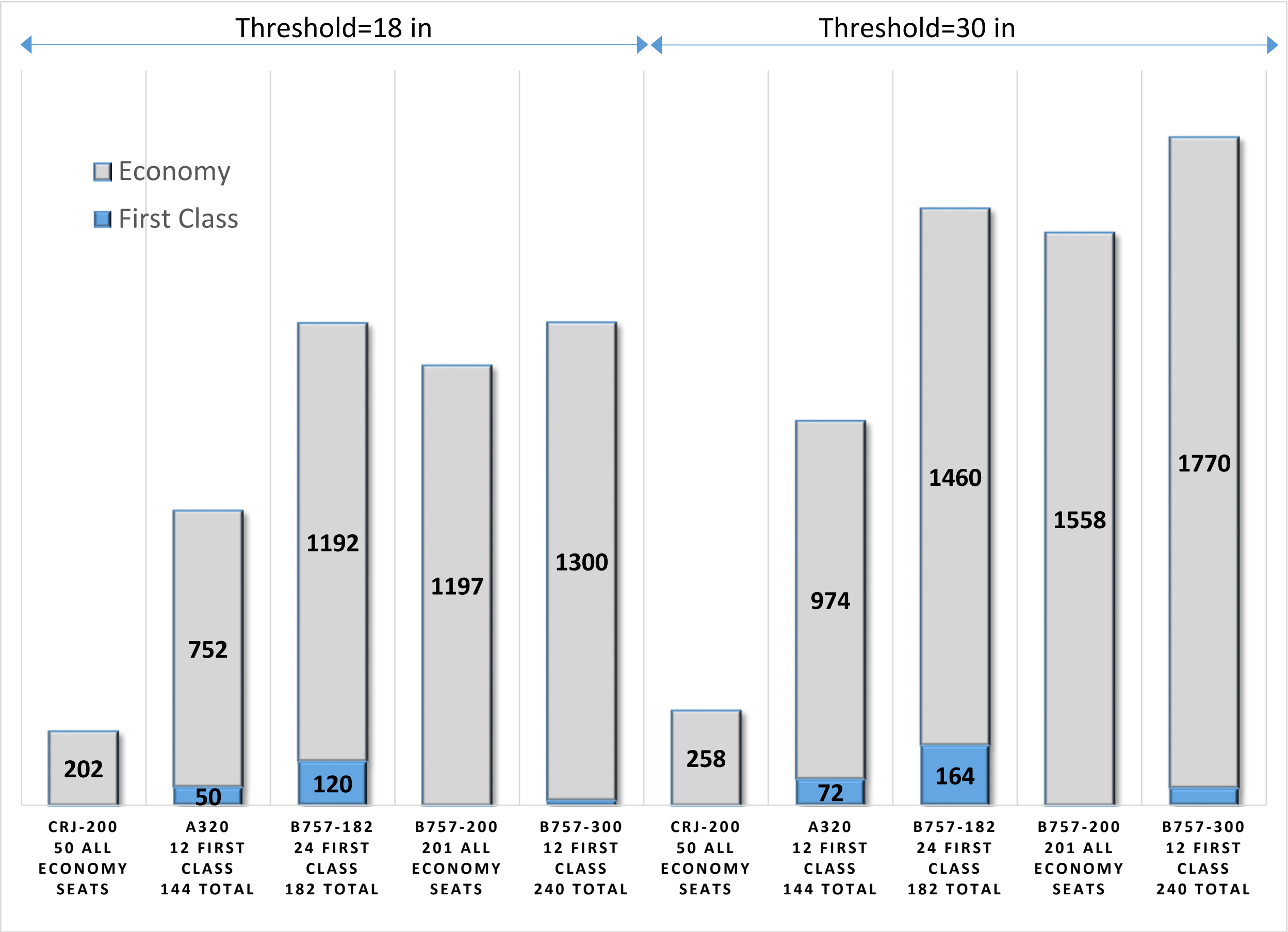

*Figure 7. Number of human-human contacts during deplaning for the five airplanes during deplaning for contact threshold of 18 inches and 30 inches*

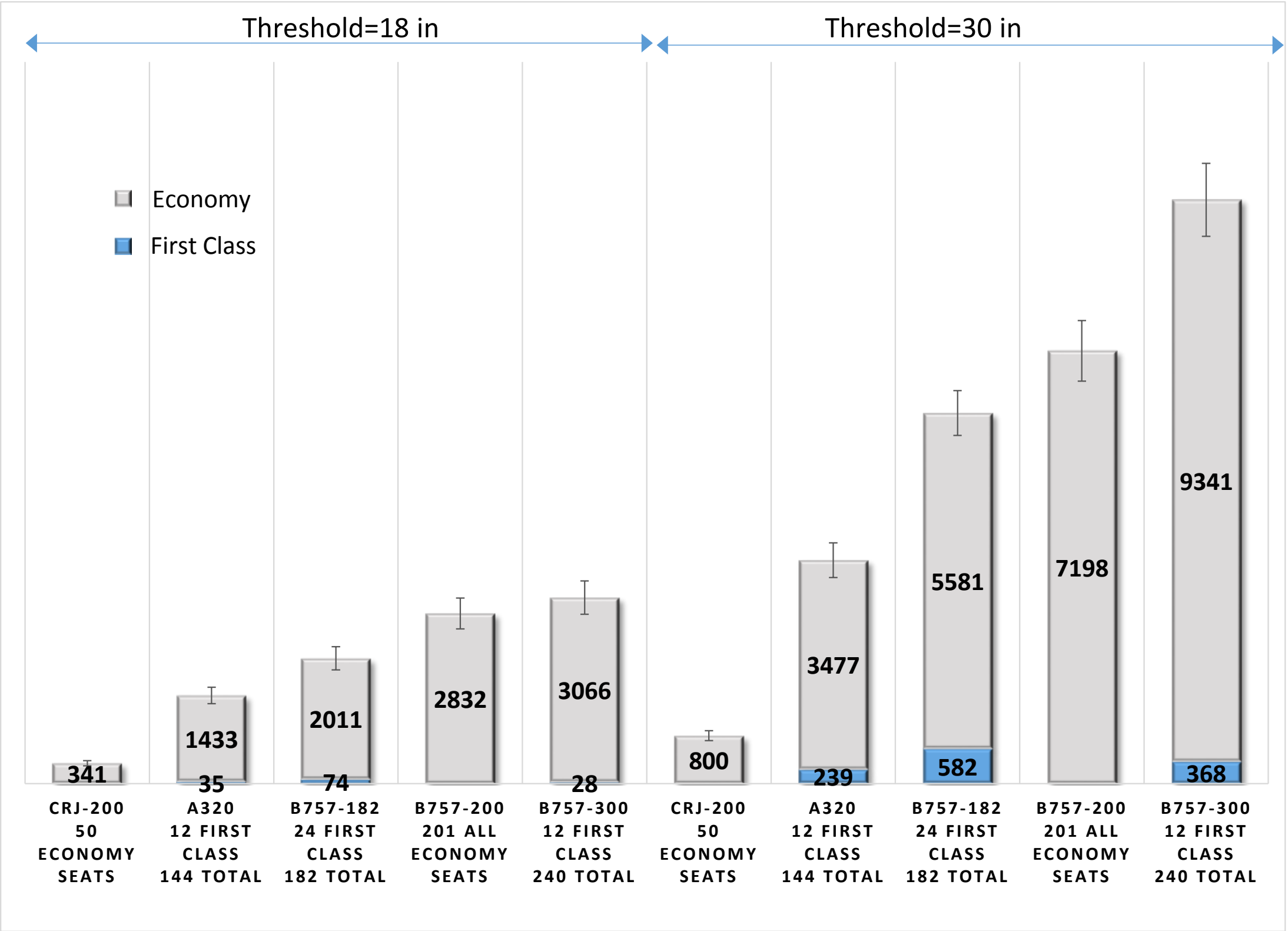

*Figure 8. Number of human-human contacts during boarding for the five airplanes during deplaning for contact threshold of 18 inches and 30 inches. The bars represent standard deviation.*

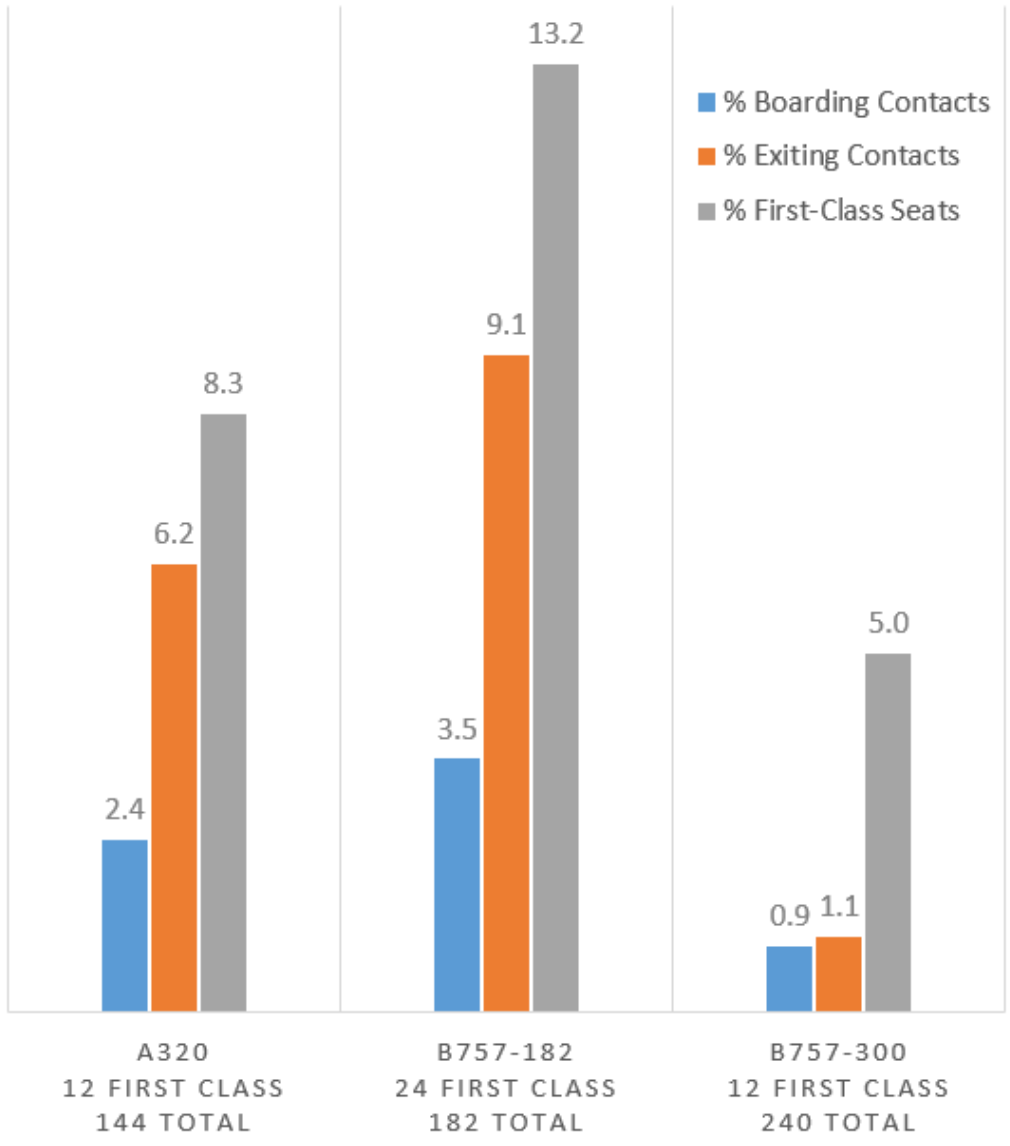

*Figure 9. Percentage of first class seats vs contacts during boarding and deplaning for contact threshold of 18 inches.*

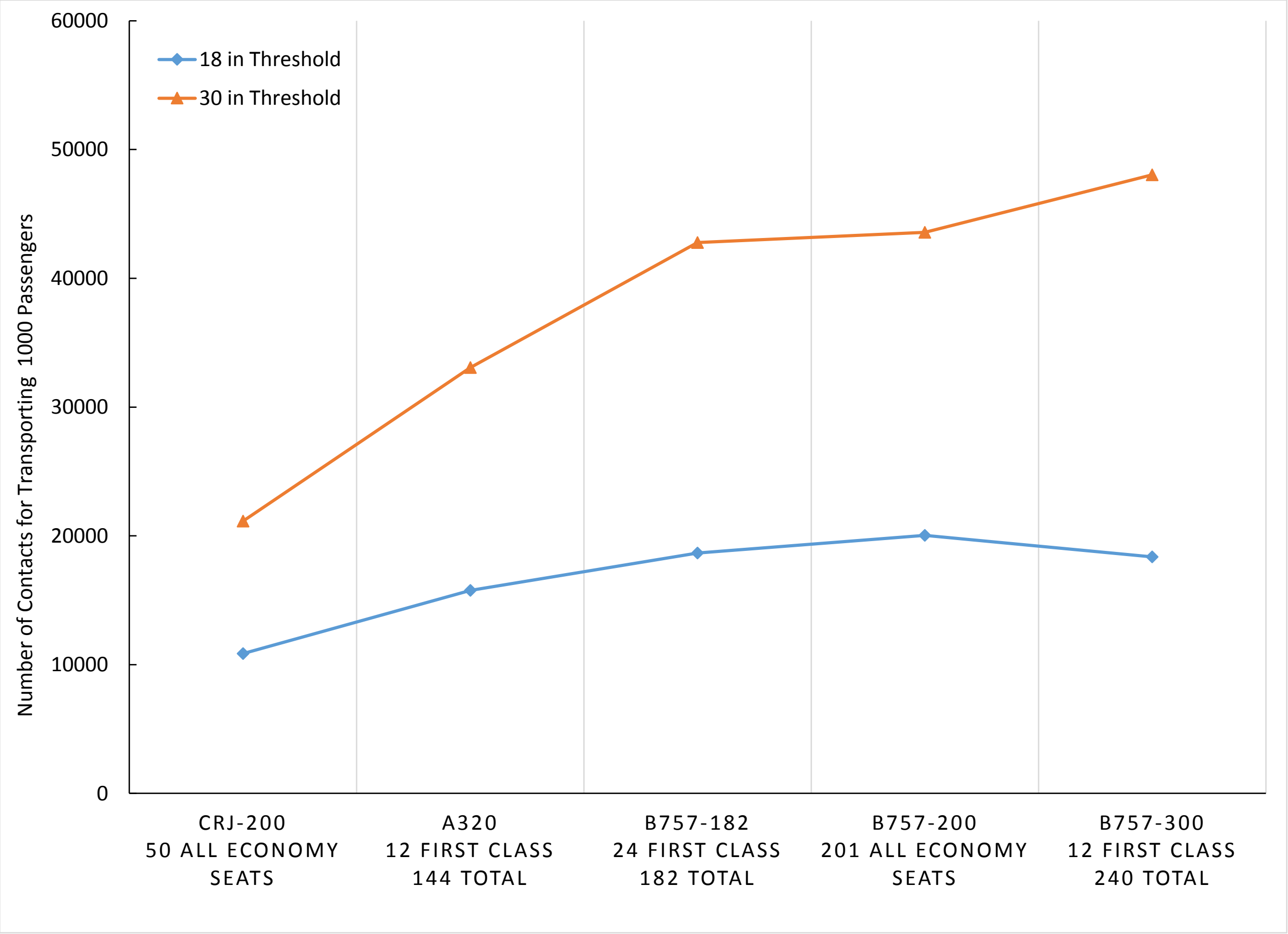

*Figure 10. Number of contacts for transporting 1000 passengers in different airplanes boarding and deplaning by default methods.*

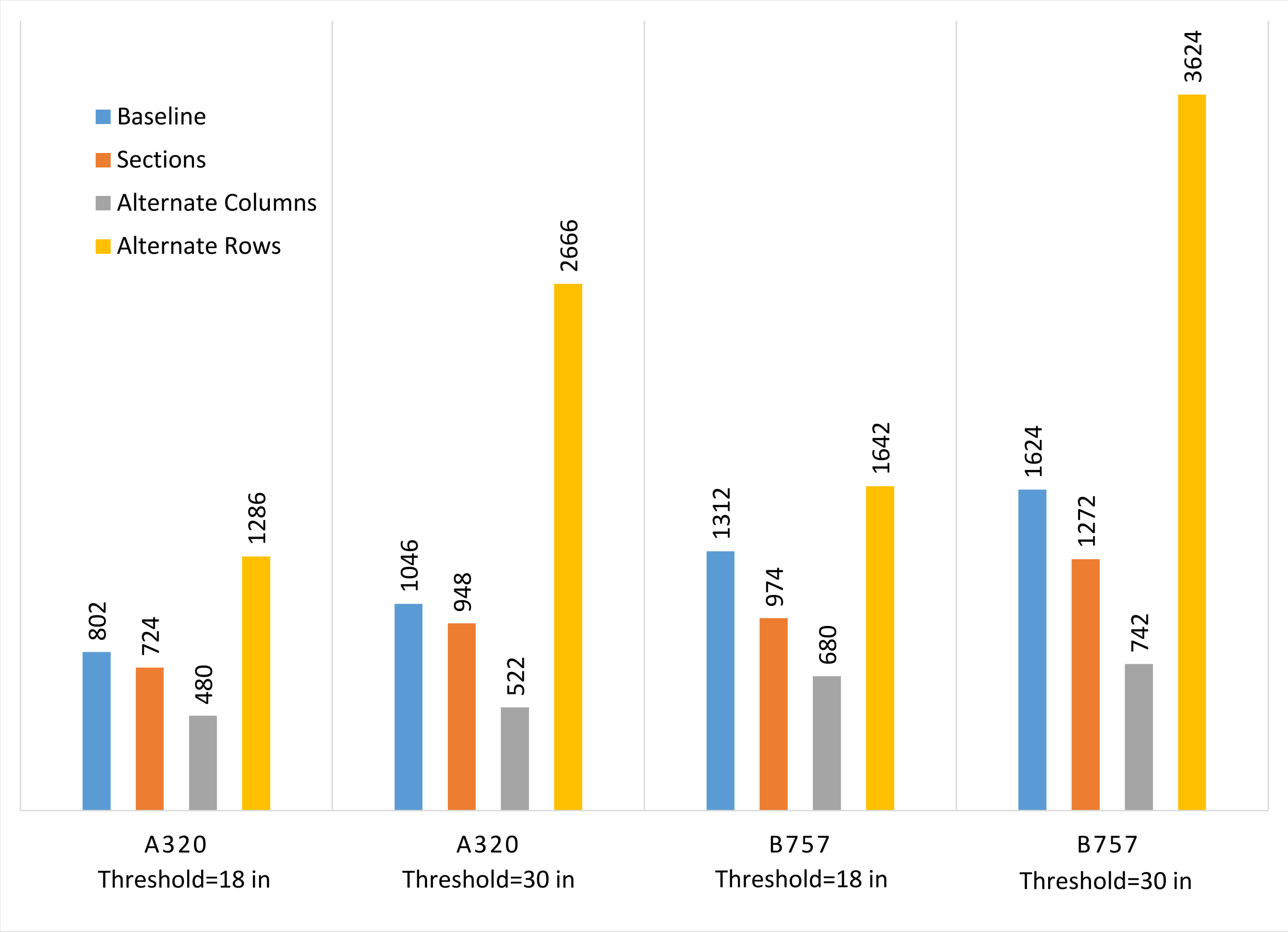

*Figure 11. Number of contacts for different deplaning strategies in 144 seat Airbus A320 and 182 seat Boeing 757-200 seating configurations.*

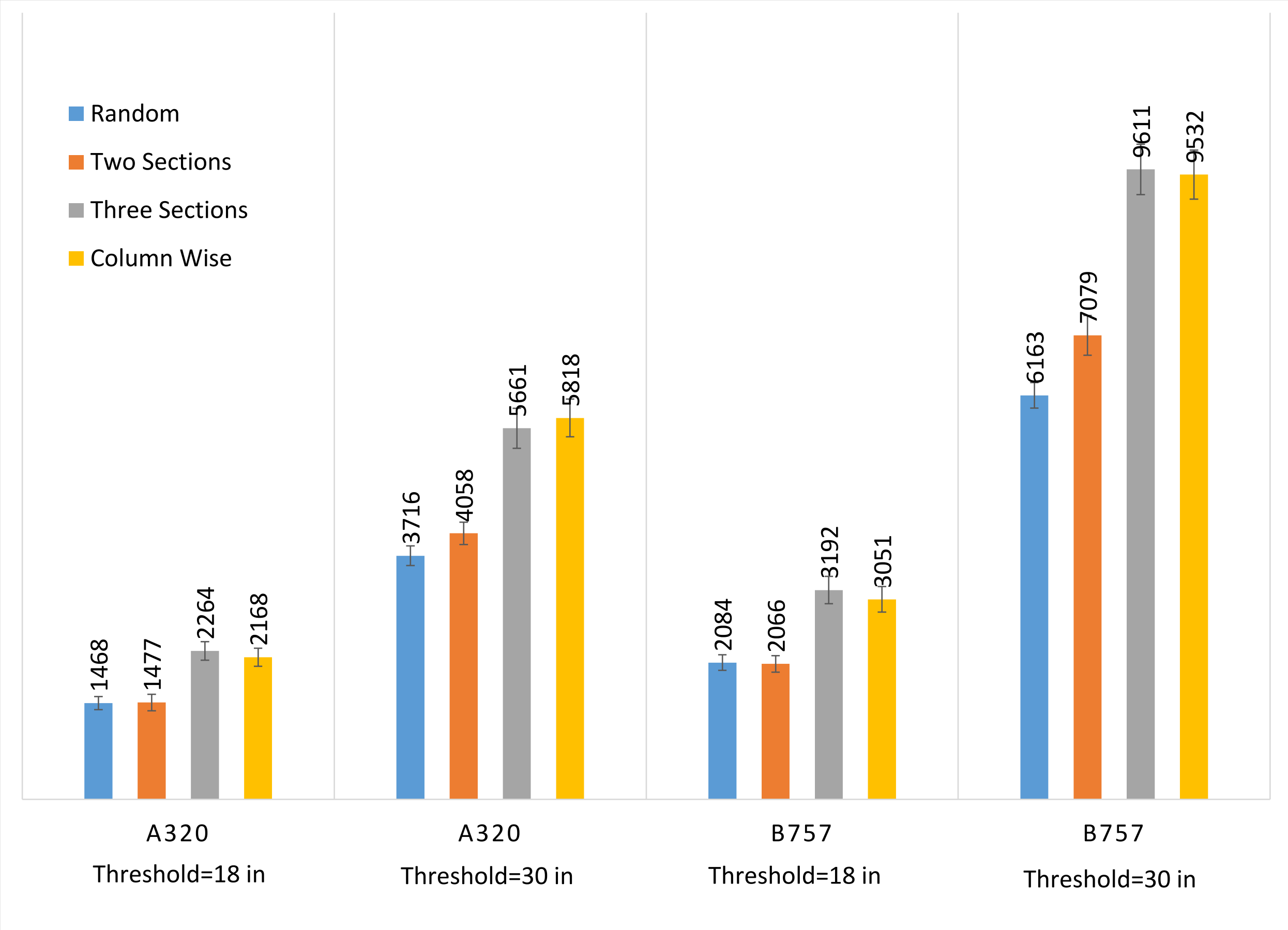

*Figure 12. Number of contacts for different boarding strategies in 144 seat Airbus A320 and 182 seat Boeing 757-200 seating configurations. The bars represent standard deviation.*